\documentclass[twocolumn,showpacs,preprintnumbers,amsmath,amssymb,aps,prl,superscriptaddress]{revtex4-1}
\usepackage{color}
\usepackage{graphicx}
\usepackage{textcomp} 
\usepackage{subeqnarray}
\usepackage[caption=false]{subfig} 
\usepackage{ulem}

\usepackage{graphicx}
\newcommand{\thetamean}{\bar{\theta}}

\begin{document}

\title{Dynamics of anchored oscillating nanomenisci}
\author{Caroline Mortagne}
\affiliation{CEMES-CNRS, UPR 8011, 29 rue Jeanne Marvig, 31055 Toulouse Cedex 4, France}
\affiliation{IMFT - Universit\'e de Toulouse, CNRS-INPT-UPS, UMR 5502, 1 all\'ee du Professeur Camille Soula, 31400 Toulouse, France}

\author{Kevin Lippera}
\affiliation{CEMES-CNRS, UPR 8011, 29 rue Jeanne Marvig, 31055 Toulouse Cedex 4, France}
\affiliation{LadHyX - UMR CNRS 7646, \'Ecole Polytechnique, Boulevard des Mar\'echaux, 91120 Palaiseau, France}

\author{Philippe Tordjeman}
\affiliation{IMFT - Universit\'e de Toulouse, CNRS-INPT-UPS, UMR 5502, 1 all\'ee du Professeur Camille Soula, 31400 Toulouse, France}

\author{Michael Benzaquen}
\affiliation{LadHyX - UMR CNRS 7646, \'Ecole Polytechnique, Boulevard des Mar\'echaux, 91120 Palaiseau, France}

\author{Thierry Ondar\c{c}uhu}
\affiliation{CEMES-CNRS, UPR 8011, 29 rue Jeanne Marvig, 31055 Toulouse Cedex 4, France}
\email{ondar@cemes.fr}

\date{\today}
\begin{abstract}
We present a self-contained study of the dynamics of oscillating nanomenisci anchored on nanometric topographical defects around a cylindrical nanofiber -- radius below 100 nm. Using frequency-modulation atomic force microscopy (FM-AFM), we show that the friction coefficient surges as the contact angle is decreased. We propose a theoretical model within the lubrification approximation that reproduces the experimental data and provides   a comprehensive description of the dynamics of the nanomeniscus. The dissipation pattern in the vicinity of the contact line and the anchoring properties are discussed as a function of liquid and surface properties in addition to the sollicitation conditions.
\end{abstract}
\maketitle

\begin{figure}[t!]
\includegraphics[scale=.9]{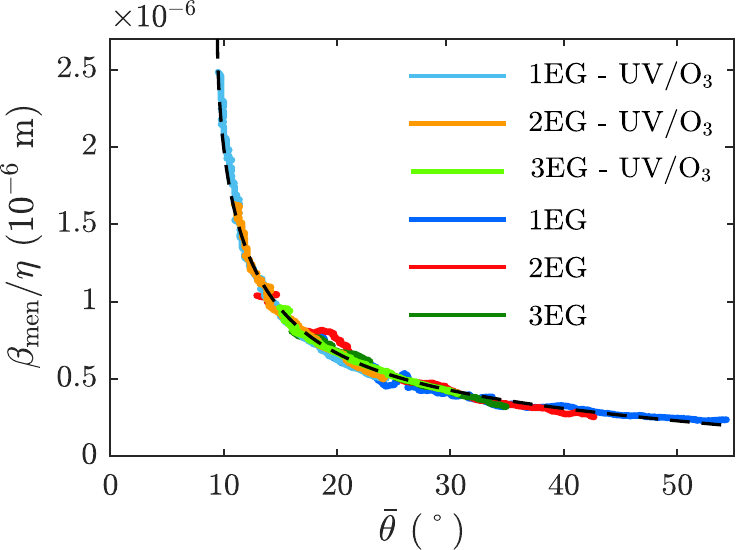} 
\caption{Normalized friction coefficient $\beta_\mathrm{men}/\eta$ plotted as a function of $\thetamean$ [see Eq.~\eqref{bmen}].  The dashed line signifies the theoretical model, while the experimental dotted curves are performed over all the studied liquids, before and after UV/O$_3$ treatment, with $R$=85~nm, $A$=18~nm and $r_0$=40~nm.}
 \label{master_curve}
\end{figure}

The study of liquid dynamics in the close vicinity of the contact line is fundamental to understand the physics of  wetting \citep{PGGRevModPhys, bonn2009}. The strong confinement inherent to this region leads, in the case of a moving contact line, to a divergence of the energy dissipation. This singularity can be released by the introduction of microscopic models based on long range interactions, wall slippage or diffuse interface \citep{snoeijerARFM2013} which are still difficult to determine experimentally. In most cases, the spreading is also controlled by the pinning of the contact line on surface defects \citep{joannydeGennes1984, PerrinPRL2016}. For nanometric defects, the intensity and localisation of the viscous energy dissipation is an open issue to understand the wetting dynamics. The aim of this paper is to study the hydrodynamics of a nanomeniscus anchored on nanometric topographic defects and subjected to an external periodic forcing. In addition to wetting dynamics on rough surfaces, this issue is relevant for vibrated droplets or bubbles \citep{Noblin2004} and for the reflection of capillary waves on a solid wall \citep{PRLFauve}.\\
Atomic force microscopy (AFM) has proven to be a unique tool to carry out measurements on liquids down to the nanometer scale: liquid structuration \citep{fukuma2010} or slippage \citep{maali2008} at solid interfaces were evidenced, while the use of specific tips fitted with either micro- or nano- cylinders  allowed quantitative measurements in viscous boundary layers \citep{PRFDupre2016} and at the contact line \citep{TongPRL2013}. In this study, we have developped an AFM experiment based on the  Frequency Modulation mode (FM-AFM) to monitor, simultaneously, the mean force and the energy dissipation experienced by an anchored nanomeniscus. Artificial defects with adjustable size are deposited on cylindrical fibers (radius~100 nm) to control the pinning of the contact line and the meniscus stretching during the oscillation. The experiments are analyzed in the frame of a nanohydrodynamics model based on the lubrification approximation. Interestingly, the meniscus oscillation does not lead to any stress divergence at the contact line allowing a full resolution without the use of cutoff lengths. 
This study thus provides a comprehensive description of dissipation mechanisms in highly confined menisci and an estimate of the critical depinning contact angle for nanometric defects.\\
The fibres used in the experimental study were carved with a dual beam FIB (1540 XB Cross Beam, Zeiss) from conventional silicon AFM probes (OLTESPA, Bruker). Using a beam of Ga ions, a 2 to 3-~$\mu$m long cylinder of radius $R\sim$ 80~nm is milled  at the end of a classical AFM tip. 
An ELPHY MultiBeam (Raith) device allows to manufacture nanometric spots
 of platinum by Electron Beam Induced
\begin{figure}[h!]
\includegraphics[scale=0.9]{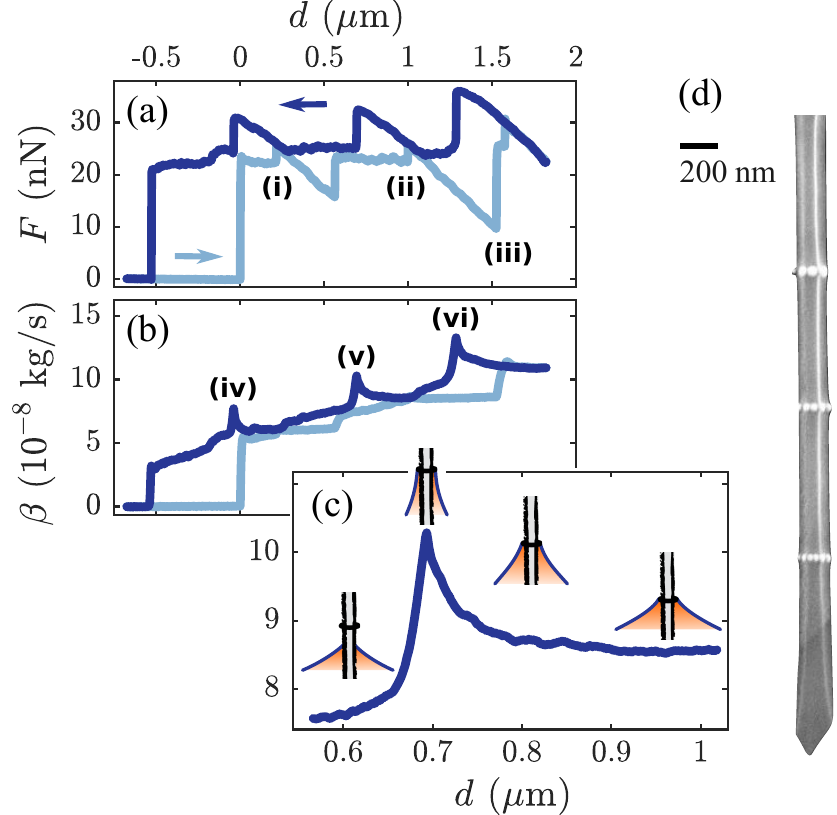}  
\caption{FM-AFM spectroscopy curves performed on a 3EG liquid drop. (a) Force $F$ and (b) friction coefficient $\beta$ as a function of the immersion depth $d$. (c) Zoom on the friction coefficient curve on the second defect with sketches of the meniscus. (d) SEM image of the 3.2~$\mu$m long and 170~nm diameter probe, covered by three platinum rings of thicknesses $r_0 = 10 ,\ 15 $ and $40$~nm, from bottom to top respectively.}
\label{spectro}
\end{figure}
Deposition (EBID) in order to create ring defect of controlled thickness around the cylinders (see Supplemental Material SM). An example of a home-made cylinder with three annular rings is displayed in Fig.~\ref{spectro}.(d). The liquids used are ethylene glycol (1EG), diethylene glycol (2EG),  triethylene glycol (3EG) and an ionic liquid, namely 1-ethyl-3-methylimidazolium tetrafluoroborate (IL). The liquids have a low volatility at room temperature. Their dynamic viscosities are $\eta =$ 19.5, 34.5, 46.5 and 44~mPa.s and their surface tensions are $\gamma=$ 49.5, 49.5, 48 and 56~mN.m at 20\textcelsius, respectively.
 As surface conditions play a crucial role in wetting,  measurements are made before and after a five minutes UV/O$_3$ treatment aimed at removing contaminants and making the surface more hydrophilic \cite{Vig1985UV}.\\
Using a PicoForce AFM (Bruker), the tips are dipped in and withdrawn from a millimetric liquid drop deposited on a silicon substrate. The experiments are performed in Frequency Modulation (FM-AFM) mode using a phase-lock loop device (HF2LI, Zurich Instrument) which oscillates the cantilever at its resonance frequency. A PID controller is used to maintain the oscillation amplitude $A$ constant. The excitation signal $A_\mathrm{ex}$ is linearly related to the friction coefficient of the interaction \cite{Giessibl2003} through $\beta = \beta_0 \ \left(A_{\mathrm{ex}}/A_{\mathrm{ex},0} - 1 \right)$, where  $A_{\mathrm{ex},0}$ and $\beta_0$ are respectively the excitation signal and the friction coefficient of the free system in air. Since we used cantilevers with high quality factor $Q \sim 200$ the resonant frequency coincides with the natural angular frequency $\omega_0= 2 \pi f$, and thus  $\beta_0 = k/(\omega_0 Q)$ where $k$ is the cantilever stiffness. The force is obtained as $F = k \bar{\delta} $ where $ \bar{\delta} $ is the mean deflection during the oscillation. \\
Figure~\ref{spectro} shows the results of a typical experiment performed on a 3EG drop. The measured force $F$ [Fig.~\ref{spectro}.(a)] and friction coefficient $\beta$ [Fig.~\ref{spectro}.(b)] are plotted as a function of the immersion depth $d$ for a ramp of 2.5~$\mu$m. The cylinder is dipped in (light blue curves) and withdrawn (dark blue curves)  from the liquid bath at 2.5~$\mu$m.s$^{-1}$.  The tip oscillates at its resonance frequency (66 820~Hz in air) with an amplitude of 6~nm. The cantilever stiffness is $k = 1.5$~N.m$^{-1}$, soft enough to perform deflection measurements while being adapted for the dynamic mode.
The force curve can be interpreted using the expression of the capillary force \citep{PRLDelmas2011}: $F= 2 \pi R \gamma \cos \thetamean $, where $R$ is the fiber radius and  $\thetamean$ is the mean contact angle during the oscillation. 
After the meniscus formation at $d$ = 0, and until the contact line anchors on the first ring (at reference (i)) $F$ and $\thetamean$ remain constant, consistent with \cite{PRLBarber2004, LangYazdanpanah2008,PRLDelmas2011}. A small jump of the force is observed when the contact line reaches a platinum ring on reference points (i), (ii) or (iii). Once the meniscus is pinned, the contact angle increases as the cylinder goes deeper into the liquid, leading to a decrease of the force $F$. Conversely, the withdrawal leads to a decrease of $\thetamean$ and an increase of the force $F$  on the left of (i), (ii) and (iii). Hence, each ring induces two hysteresis cycles characteristic of strong topographic defects \citep{joannydeGennes1984}.\\
\begin{figure}[t!]
\includegraphics[scale=0.47]{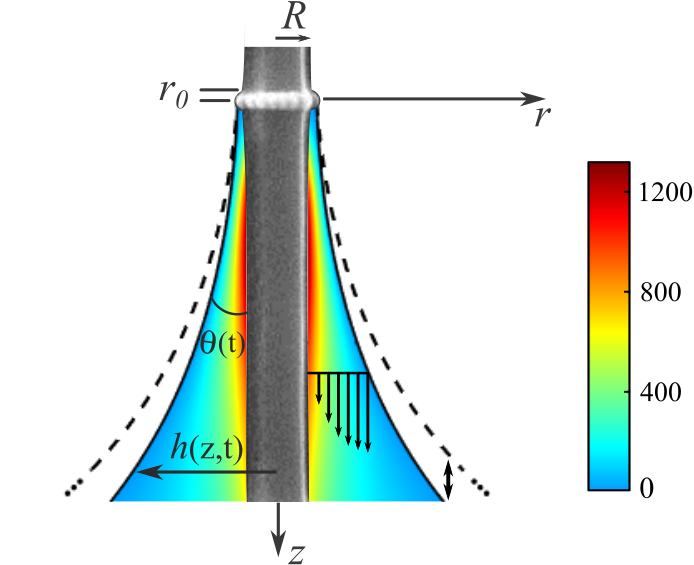}  
\caption{Oscillating meniscus anchored on a defect, displayed in the frame of reference of the fibre. The velocity profile (black arrows) is calculated from Eq.~\eqref{v}. The stress field  $\eta \partial _r v$ (color gradient) is computed for $R$=100~nm, $r_0$ = 30~nm, $l_c$=2~mm, $A$=10~nm, $f=65$~kHz, $
\theta_e$ = 10\textdegree and $\eta = 30~$mPa.s. Colorbar in Pa.}
\label{schema}
\vspace{-0.2 cm}
\end{figure}
\noindent Different contributions to the probe-liquid system account for the friction coefficient behavior. The global increase of $\beta$ with $d$ observed on Fig.~\ref{spectro}(b) results from the contribution of the viscous layer around the tip which is proportional to the immersion depth \citep{PRFDupre2016}. At withdrawal, $\beta$ increases dramatically when the probe reaches the reference points (iv), (v) and (vi) of Fig.~\ref{spectro}(b). In those regions, the force curve indicates that the meniscus is pinned on a defect. The dissipation growth is therefore attributed to the decrease of the contact angle before depinning as schematized on the zoom on the friction coefficient curve [Fig. \ref{spectro}(c)]. This large effect can be qualitatively understood  considering that small contact angles -- corresponding to reduced film thickness --  generate strong velocity gradients in the meniscus and thus a large dissipation. Note that a similar behaviour is observed on a moving contact line for which the friction coefficient also displays a stong dependance upon the contact angle $\beta \sim 1/\thetamean$ \citep{PGGRevModPhys}.\\
In order to account for the experimental results, we developed a theoretical model for the oscillation of a liquid meniscus in cylindrical geometry (see SM). We consider the problem in the frame of reference attached to the cylinder (see Fig.~\ref{schema}).
 The flow induced by the interface motion leads to a friction coefficient $\beta_{\mathrm{men}}$. The latter is  related to the mean energy loss $\mathcal{P}$ during an oscillation cycle, through  \mbox{$\mathcal{P}= \beta_\mathrm{men} (A\omega)^2/2 $} \citep{perez2001book}. Since the capillary number is small -- \mbox{$Ca = A \omega \eta/\gamma \sim 10^{-3}$} -- we may safely state that viscous effects do not affect the shape of the liquid interface. Therefore, the meniscus profile is solution of the Laplace equation resulting from the balance between  capillary and hydrostatic pressures, which in turn yields the well known catenary shape \citep{derjaguin1946theory,JFMJames1974, dupreLangmuir2015}:
\begin{equation}
h =  (R + r_0)\cos\theta \cosh\left( \frac{z}{(R + r_0) \cos\theta} -  \text{ln}(\zeta)\right) \ ,
\end{equation} 
were $\zeta=\cos\theta/[1 + \sin\theta]$.
The meniscus height $Z_0$ is given, in the limit of small contact angles, by: 
\begin{equation} 
Z_{0} = ( R + r_0)\cos\theta \left[ \text{ln} \left(\dfrac{4\  l_{\text{c}}}{R+r_0}\right) - \gamma_E \right] \ ,
\label{Z_0}
\end{equation} 
 with $\gamma_E\simeq 0.577$ the Euler constant and $l_c$ the capillary length. Since $Z_0(t)$ oscillates around its mean position as \mbox{$Z_{0}(\theta(t)) = Z_0(\thetamean) + A \cos(\omega t)$}, we can derive the temporal evolution of the contact angle:
\begin{equation}
\cos \theta (t)  =  \cos \thetamean  +  \frac{A \cos (\omega t) }{ (R + r_0)\left[ \text{ln} \left(\frac{4 l_{c}}{R + r_0}\right) - \gamma_E \right]} \ .
\label{theta_t}
\end{equation}
Note that our model is meant to deal with positive contact angles only, even if the defect thickness could in principle allow slightly negative ones. This defines a critical contact angle $\theta_\mathrm{crit}$ related to the minimum value of $\thetamean$ allowed by the model. One has: 
 {$\cos\theta_\mathrm{crit}  = 1-   A/\left[ (R + r_0) \left(\text{ln} \left({4\  l_{c}}/({R + r_0})\right) - \gamma_E \right)\right] $}.  This critical depinning angle on an ideally strong defect increases with respect to $A$ and decreases with respect to $R + r_0$.
The interface motion being known, the velocity field is derived using the Stokes equation. Indeed, gravity and inertia can be safely  neglected ($Re \sim 10^{-8} $ and $l_c \simeq $ 2~mm). Moreover, the  viscous diffusion timescale $\tau_\nu = R^2/\nu$ is much smaller than the oscillation period ($\tau_\nu f \sim 10^{-7}$), such that the Stokes equation reduces to the simplest steady Stokes equation. Using the lubrication approximation, we have finally $\partial_z P  \ = \  \eta \Delta_r v$ where $P$ is the hydrodynamic pressure and $v$ is the velocity component in the $z$ direction. 
 Finally, combining  the mass conservation equation -- \mbox{$\partial_t   (\pi h^2 ) + \partial_z q = 0$} -- 
where $q$ is the local flow rate through a liquid section of normal $z$, the no-slip  (at  $r = R$) and free interface (at  $r = h$) boundary conditions, yields the velocity profile:
 \begin{equation}
v (r,z,t)=  \dfrac{ 2  \left[ R^2 +2\ h^2 \ \text{ln}\left( {r}/{R}\right) - r^2 \right] \textstyle{\int_{0}^{z}}\text{d}u\, {\partial_t(h^2)}   }{ R^4 + 3\ h^4 - 4\ h^2 \ R^2 - 4\ h^4  \text{ln}\left({h}/{R}\right) }\ .
\label{v}
\end{equation}
From Eq.~\eqref{v} we derive the expression of $\beta_{\mathrm{men}}$:
\begin{equation}
\beta_{\mathrm{men}}(\thetamean) =  \left< \frac{4 \pi \eta}{A^2 \omega^2} \int_{0}^{Z_0} \int_{R}^{h}  \ \left( \partial_r v \right)^2  r  \mathrm{d}r \mathrm{d}z \right>_t \ ,
\label{bmen}
\end{equation}
where $<\ >_t$, designates the temporal average over an oscillation cycle (see SM).
Figure~\ref{schema} displays an example of viscous stress field (color gradient) and velocity profile (vertical dark arrows) inside a nanomeniscus pinned on a defect with $r_0$=40~nm, for typical operating conditions. We observe that the stress is essentially localized at the fiber wall and strongly decays when $z$ becomes of the order of a few probe radii. Hence, the lubrication approximation -- only valid for small depths and small surface gradients ($\partial_z h  \ll 1$) -- is strengthened. When the mean contact angle $\thetamean$ is decreased a strong increase of the viscous stress is observed but its localization remains mostly unchanged (see SM). A striking result is the influence of the defect size. For contact angles close to the critical one, reduction in size of the defect increases significantly the viscous stress in a region closer to the contact line (see SM).
Figure~\ref{master_curve} displays an example of normalized friction coefficient curve  $\beta_\mathrm{men}/\eta$ (dashed line), plotted as  function of $\thetamean$. A significant increase of $\beta_\mathrm{men}$ is observed for decreasing contact angles in agreement with the experimental observations.\\ 
\noindent To quantitatively confront the FM-AFM experiments to the theoretical model, we use the force signal to determine the experimental contact angles $\thetamean$. We assume that, due to the inhomogeneous thickness of the platinum rings, the meniscus depins from the defect for a contact angle $\theta_\mathrm{break}$ larger than $\theta_\mathrm{crit}$ value expected for an ideal defect. The maximum force before depinning then reads \mbox{$F_\mathrm{max} = 2 \pi \gamma (R + r_0)  \cos \theta_\mathrm{break} $} which allows to calculate the experimental contact angle for any $d$ values using 
\mbox{${\cos\thetamean} = {[F/F_\mathrm{max}]}\,{\cos \theta_\mathrm{break}}$.
}
The latter equation enables to determine the contact angle for each $d$ position without using the cantilever stiffness only known within 20 \% error.
For each experiment, we make a linear fit of the whole friction coefficient curve without taking into account the regions influenced by the defects. The subtraction of this fit allows to dispose of the viscous layer contribution, leaving only $\beta_\mathrm{men}$ and a constant term induced by the bottom of the tip, called $\beta_\mathrm{bottom}$. The data are then fitted by computing the parameters $\beta_\mathrm{bottom}$ and  $\theta_\mathrm{break}$ which minimise the standard deviation between the experimental data and the theoretical curve [Eq.~\eqref{bmen}]. As for $R$ and $r_0$, we use effective values measured by SEM. FM experiments were then performed over all the studied liquids. More than ninety experiments were carried out with two different home-made probes (R = 80~nm  and 85~nm),  defect thicknesses $r_0$ between 10 and 50~nm and oscillation amplitudes $A$ ranging from 5 to 35~nm. Additionally, experiments were performed before and after surface cleaning by UV/O$_3$ treatment to assess the influence of tip wettability.\\
As an example, Fig.~\ref{master_curve} displays six curves performed with three different liquids, before and after UV/O$_3$ treatment, on the same defect ($R$=85~nm and $r_0$=40~nm) with an amplitude $A$=18~nm. 
The agreement between the experimental data and the theoretical model is remarkable.
A ten-fold enhancement of dissipation is observed when the contact angle is decreased from 50\char23 to 10\char23. As expected, the five minute surface cleaning does not affect the dissipation process since all curves superpose on a same master curve. Yet, ozone cleaning has a strong impact on the $\theta_\mathrm{break}$ values. The hydrophilic surfaces obtained after UV/O$_3$ treatment lead to a strong pinning which allows to reach smaller contact angle values. For example, for 1EG $\theta_\mathrm{break}$ decreases from 18.5\char23 to 9.5\char23, the latter value being very close from the value of $\theta_\mathrm{crit} = 9,4$\char23.  Consequently, the dissipation can reach larger values after ozone treatment. This is a common observation on all the measurements. When the tip is more hydrophobic, the liquid may detach between the dots forming the defect before the $\theta_\mathrm{crit}$ value is reached.
\begin{figure}[t!]
\includegraphics[scale=1]{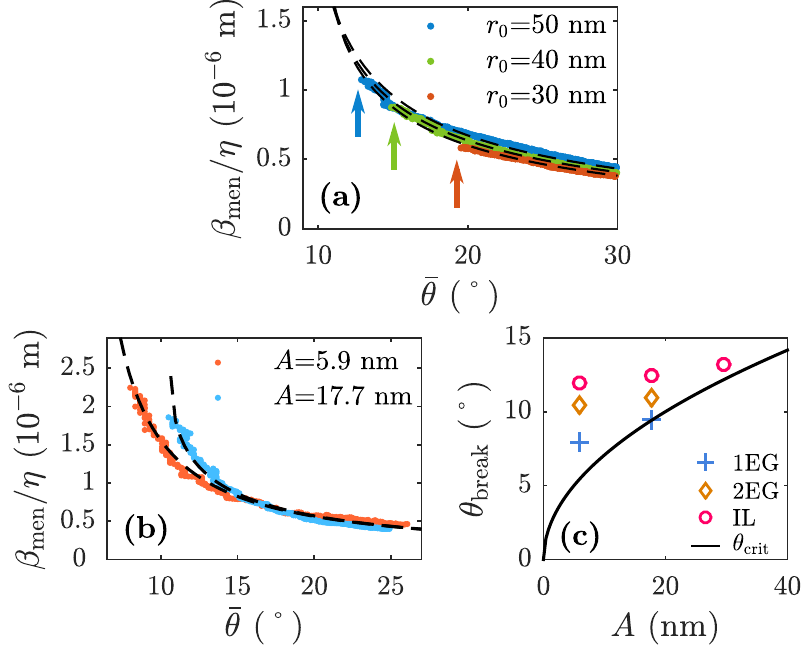}  
\caption{(Normalized friction coefficient $\beta_\mathrm{men}/\eta$ vs mean contact angle $\thetamean$ for different operating conditions. The dashed lines are plots of the theoretical model  [Eq.~\eqref{bmen}]. (a) Influence of ring thickness $r_0$. The arrows indicate the value of $\theta_\mathrm{break}$. (b) Influence of oscillation amplitude $A$. (c) Plot of $\theta_\mathrm{break}$(symbols) and $\theta_\mathrm{crit}$ (solid line) as a function of the oscillation amplitude for a defect of thickness $r_0 $= 40 nm.}
 \label{O3-r0_A}
\end{figure}
\noindent In order to discuss further the influence of the various parameters and the resulting values of the fitting variables $\theta_\mathrm{break}$ and $\beta_\mathrm{bottom}$, we reported on Figure~\ref{O3-r0_A} a comparison between the theoretical model and FM experiments performed on 3EG for (a) different defect thicknesses  and (b) various oscillation amplitudes.
Figure~\ref{O3-r0_A}(a) shows that the ring thickness $r_0$ has a low impact on the friction coefficient curve for $30~\textrm{nm} \leq r_0 \leq 50~\textrm{nm}$. Nevertheless, a systematic evolution of $\theta_\mathrm{break}$ is observed: larger defect thicknesses lead to a stronger pinning of the defect which results in a smaller $\theta_\mathrm{break}$ value, as marked by the arrows on the curves. We also found that the oscillation amplitude only plays a significant role for contact angles close to $\theta_\mathrm{crit}$. Therefore its influence can only be noticed after the UV/O$_3$ treatment. The theoretical model reproduces well the influence of amplitude observed for contact angles smaller than 15\char23  [see Fig.~\ref{O3-r0_A}(b)]. A larger amplitude increases slightly the value $\beta_\mathrm{men}$ at low $\thetamean$ and also leads to an increase of the $\theta_\mathrm{break}$ value,  a general trend observed on all experiments. On hydrophilic tips [see Fig. \ref{O3-r0_A}(c)], $\theta_\mathrm{break}$ approaches the $\theta_\mathrm{crit}$ value expected for an ideal defect, but dynamic effects are also probably involved since an effect of the liquid nature is observed.\\
The  limited influence of the experimental parameters on the dissipation in the meniscus justifies reporting all the experimental results on a same curve (see SM) showing a general trend  well reproduced by the model using two adjustable parameters. As expected, contrary to  $\theta_\mathrm{break}$, $\beta_\mathrm{bottom}$ does not show any systematic influence of amplitude, defect size and wettability. Statistics over all experiments (see histogram in SM) show that $\beta_\mathrm{bottom}$ is proportional to the liquid viscosity and lead to $\beta_\mathrm{bottom}/(\eta R) = 7 \pm 3.5$. If we assimilate the cylinder bottom to a disk of radius $R$, the dissipation induced by the fibre bottom is given by  $\beta_\mathrm{bottom}= 8 \eta\ R$ (see ref.~\citep{stone1998}), consistent with the experimental results. However, quantitative comparison with the theory is compromised due to the ill-defined shape of the tip end.\\
In conclusion, the development of dedicated AFM probes with defects of controlled size down to nanometer scale, combined with the use of frequency-modulation AFM, enables the accurate investigation of the viscous dissipation in anchored oscillating menisci.  We find an excellent agreement between the experimental results and our lubrication based theoretical model describing the flow pattern inside the oscillating meniscus. The stretching of the meniscus leads to a strong increase of viscous stress which accounts for the surge of dissipated energy observed at small angle. Note that this effect is amplified for small defect sizes, in which case the stress is strongly localised at the contact line with important consequences on the wetting dynamics on surfaces with defects. Our results also give new insights on the depinning of the contact line from defects which appears for a contact angle value $\theta_\mathrm{break}$ larger than the theoretical one $\theta_\mathrm{crit}$ obtained for a perfect pinning. The latter value could be approached using hydrophilic tips showing that the pinning is all the stronger that the oscillation amplitude $A$ is small and the defect size $r_0$ is large.
This study demonstrates that FM-AFM is a unique tool for quantitative measurements of dissipation in confined liquids, down to the nanometer scale, and paves the way for a systematic study of open questions in wetting science regarding the extra dissipation which occurs when the contact line starts to move.\section{Acknowledgments}
The authors thank P. Salles for his help in the development of tip fabrication procedures, Dominique Anne-Archard for viscosity measurements and J.-P. Aim\'e, D. Legendre and E. Rapha\"el  for fruitful discussions. This study has been partially supporter through the ANR by the NANOFLUIDYN project (grant n\char23 ANR-13-BS10-0009). 
\bibliographystyle{h-physrev}
\bibliography{biblio}

\begin{thebibliography}{10}

\bibitem{PGGRevModPhys}
P.~G. de~Gennes,
\newblock Rev. Mod. Phys. {\bf 57}, 827 (1985).

\bibitem{bonn2009}
D.~Bonn, J.~Eggers, J.~Indekeu, J.~Meunier, and E.~Rolley,
\newblock Rev. Mod. Phys {\bf 81}, 739 (2009).

\bibitem{snoeijerARFM2013}
J.~H. Snoeijer and B.~Andreotti,
\newblock Ann. Rev. Fluid Mech. {\bf 45}, 269 (2013).

\bibitem{joannydeGennes1984}
J.~Joanny and P.-G. De~Gennes,
\newblock J. Chem. Phys. {\bf 81}, 552 (1984).

\bibitem{PerrinPRL2016}
H.~Perrin, R.~Lhermerout, K.~Davitt, E.~Rolley, and B.~Andreotti,
\newblock Phys. Rev. Lett. {\bf 116}, 184502 (2016).

\bibitem{Noblin2004}
X.~Noblin, A.~Buguin, and F.~Brochard-Wyart,
\newblock Eur. Phys. J. E {\bf 14}, 395 (2004).

\bibitem{PRLFauve}
G.~Michel, F.~P\'etr\'elis, and S.~Fauve,
\newblock Phys. Rev. Lett. {\bf 116}, 174301 (2016).

\bibitem{fukuma2010}
T.~Fukuma,
\newblock Science and Technology of Advanced Materials {\bf 11}, 033003 (2010).

\bibitem{maali2008}
A.~Maali, T.~Cohen-Bouhacina, and H.~Kellay,
\newblock Appl. Phys. Lett. {\bf 92}, 053101 (2008).

\bibitem{PRFDupre2016}
J.~Dupr\'e~de Baubigny {\em et~al.},
\newblock Phys. Rev. Fluids {\bf 1}, 044104 (2016).

\bibitem{TongPRL2013}
S.~Guo {\em et~al.},
\newblock Phys. Rev. Lett. {\bf 111}, 026101 (2013).

\bibitem{Vig1985UV}
J.~R. Vig,
\newblock J. Vac. Sci. Technol. A {\bf 3}, 1027 (1985).

\bibitem{Giessibl2003}
F.~J. Giessibl,
\newblock Rev. Mod. Phys. {\bf 75}, 949 (2003).

\bibitem{PRLDelmas2011}
M.~Delmas, M.~Monthioux, and T.~Ondar\ifmmode~\mbox{\c{c}}\else \c{c}\fi{}uhu,
\newblock Phys. Rev. Lett. {\bf 106}, 136102 (2011).

\bibitem{PRLBarber2004}
A.~H. Barber, S.~R. Cohen, and H.~D. Wagner,
\newblock Phys. Rev. Lett. {\bf 92}, 186103 (2004).

\bibitem{LangYazdanpanah2008}
M.~M. Yazdanpanah {\em et~al.},
\newblock Langmuir {\bf 24}, 13753 (2008).

\bibitem{perez2001book}
J.-P. P{\'e}rez,
\newblock {\em M{\'e}canique: fondements et applications: avec 300 exercices et
  probl{\`e}mes r{\'e}solus} (Dunod, 2001).

\bibitem{derjaguin1946theory}
B.~Derjaguin,
\newblock Dokl. Akad. Nauk SSSR {\bf 51}, 517 (1946).

\bibitem{JFMJames1974}
D.~F. James,
\newblock J. Fluid Mech. {\bf 63}, 657 (1974).

\bibitem{dupreLangmuir2015}
J.~Dupré~de Baubigny {\em et~al.},
\newblock Langmuir {\bf 31}, 9790 (2015).

\bibitem{stone1998}
W.~Zhang and H.~A. Stone,
\newblock J. Fluid Mech. {\bf 367}, 329 (1998).

\end{thebibliography}

\onecolumngrid

\clearpage

\includegraphics[page=1,scale=0.92]{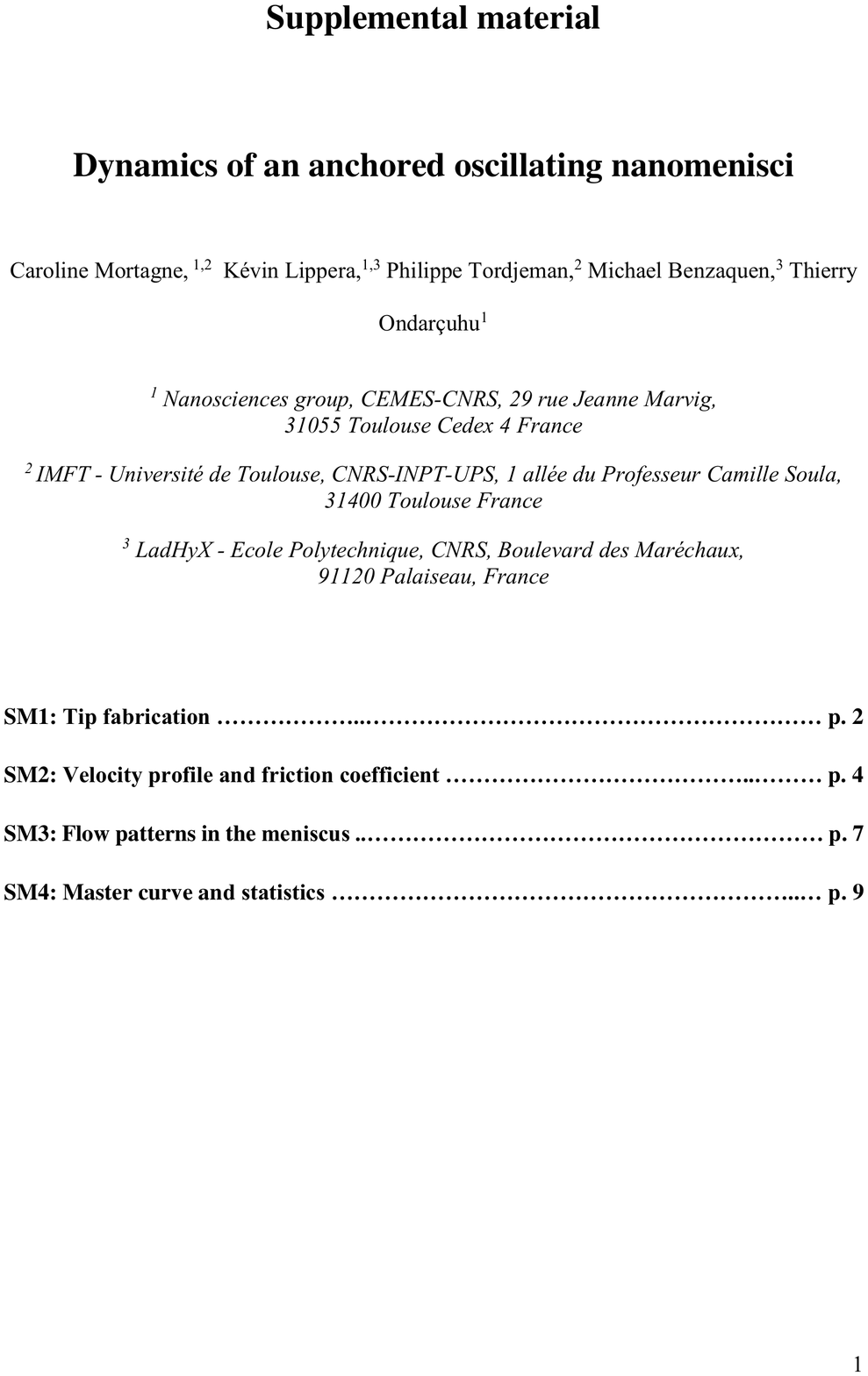}      
\clearpage
\includegraphics[page=2,scale=0.92]{SM.pdf}    
\clearpage
\includegraphics[page=3,scale=0.92]{SM.pdf}    
\clearpage
\includegraphics[page=4,scale=0.92]{SM.pdf}    
\clearpage
\includegraphics[page=5,scale=0.92]{SM.pdf}    
\clearpage
\includegraphics[page=6,scale=0.92]{SM.pdf}    
\clearpage
\includegraphics[page=7,scale=0.92]{SM.pdf}    
\clearpage
\includegraphics[page=8,scale=0.92]{SM.pdf}      
\clearpage
\includegraphics[page=9,scale=0.92]{SM.pdf}    
\clearpage
\includegraphics[page=10,scale=0.92]{SM.pdf}    

\end{document}